\documentclass[proceedings]{stacs}
\stacsheading{2010}{417-428}{Nancy, France}
\firstpageno{417}

\usepackage{graphicx}
\usepackage{cclicenses}

\newcommand{\frakS}{\mathfrak{S}}
\newcommand{\fraks}{\mathfrak{s}}
\newcommand{\Pfin}{\mathfrak{P}_{\textit{fin}}}

\newcommand{\N}{\mathbb{N}}

\newcommand{\Z}{\mathbb{Z}}

\def\[#1]{\hbox{$ [\kern -.4em [\, {#1}\, ]\kern -.4em]$}}
\newcommand{\means}[1]{\hbox{$ [\kern -.4em [\, {#1}\, ]\kern -.4em]$}}

\newcommand{\sta}{\text{sta}}

\newcommand{\ext}{\text{ext}}
\newcommand{\inte}{\text{int}}
\newcommand{\inp}{\text{input}}
\newcommand{\dyn}{\text{dyn}}
\newcommand{\ini}{\text{ini}}

\newcommand{\go}{\text{go}}
\newcommand{\acc}{\text{acc}}
\newcommand{\rej}{\text{rej}}

\newcommand{\Bool}{{\tt Bool}}

\newcommand{\suc}{\textit{Succ}}
\newcommand{\pred}{\textit{Pred}}

\newcommand{\dom}{\text{\tt Domain}}

\newcommand{\pos}{\text{pos}}
\newcommand{\adr}{\text{addr}}
\newcommand{\cast}{\text{cast}}

\newcommand{\gra}{\text{gra}}

\begin{document}

\title[Evolving MultiAlgebras unify all usual sequential computation models]{Evolving MultiAlgebras\newline
unify all usual sequential computation models}

\author[lab1]{S. Grigorieff}{Serge Grigorieff}
\address[lab1]{LIAFA, CNRS \& Universit\'e Paris Diderot - Paris 7,
Case 7014, \ 75205 Paris Cedex 13}
\email{seg@liafa.jussieu.fr}
\urladdr{http://www.liafa.jussieu.fr}

\author[lab2]{P. Valarcher}{Pierre Valarcher}
\address[lab2]{LACL, Universit\'e de Paris Est,
IUT Fontainebleau/S\'enart,
Route de Hourtault
77300 Fontainebleau}
\email{valarcher@univ-paris12.fr}
\urladdr{http://lacl.univ-paris12.fr/valarcher/}

\keywords{Abstract state machines;
Models of machines;
Computability;
Universality;
Logic in computer science;
Theory of algorithms}
\begin{abstract}
\noindent
It is well-known that Abstract State Machines (ASMs)
can simulate ``step-by-step" any type of machines
(Turing machines, RAMs, etc.).
We aim to overcome two facts:
1) simulation is not identification,
2) the ASMs simulating machines of some type do not
constitute a natural class among all ASMs.
We modify Gurevich's notion of ASM
to that of
EMA (``Evolving MultiAlgebra")
by replacing the program (which is a syntactic object)
by a semantic object:
a functional which has to be very simply definable over the
static part of the ASM.
We prove that very natural classes of EMAs
correspond via ``literal identifications"
to slight extensions of the usual machine models
and also to grammar models.
Though we modify these models,
we keep their computation approach:
only some contingencies are modified.
\\
Thus, EMAs appear as the mathematical model
unifying all kinds of sequential computation paradigms.
\end{abstract}

\maketitle
\vspace{-1cm}
\tableofcontents

\section{Introduction}\label{s:intro}
%
\noindent{\bf What we prove in this paper.}
The fact that Abstract State Machines (ASMs) can strict lock-step
(i.e. ``step-by-step
") simulate
any type of machines (Turing machines, stack automata, RAM, etc)
and grammars was shown long ago by Gurevich
\cite{gurevich91, gurevich97Schonage}.
A systematic study is also done in B\"orger \cite{borger04}.
A tighter notion of simulation is also valid as shown in
Blass, Dershowitz \& Gurevich \cite{blassdershowitzgurevich09}.
\\
The questions we consider in this paper are the following:
\begin{enumerate}\em
\item[(Q1)]
Can we replace strict lock-step simulation by literal identity
(up to a simple change of view)?
\item[(Q2)]
Given a computation model $\+C$,
is it possible to get a natural characterization of the class of ASMs
which are equivalent to machines in $\+C$?
\end{enumerate}
As far as we know, up to now, there is only one isolated answer
which is about question (Q2):
Gurevich \& al. \cite{gurevich97Schonage}
proved that Sch\"onhage Storage Modification Machines
correspond exactly  (for strict lock-step equivalence)
to ASMs with unary functions only. 
\\
We bring positive answers to both questions
for the diverse usual computation models $\+C$
(Turing machines, stack automata, RAMs, Sch\"onhage Machines,
Chomsky type $0$ grammars, etc.) slightly extended to models $\+C^+$
using a tailored version of ASMs which
(resurrecting Gurevich's original name for ASMs)
we call {\em Evolving Multialgebras} (EMAs).
These answers have the following remarkably simple form:
\begin{theorem}\label{thm:main}
There exists a family of EMA static parts $\+M$
(fixed semantical feature)
and a family of dynamic signatures $\+S$
(fixed syntactical feature) such that,
letting $\+E_{M,S}$ be the family of EMAs
with static part in $\+M$ and dynamic signature in $\+S$,
\\-\ any computation device in $\+C^+$ is literally identical to some
EMA in $\+E_{\+M,\+S}$,
\\-\ this ``literal identity" correspondence is a bijection
from $\+C^+$ onto $\+E_{\+M,\+S}$.
\end{theorem}
Of course, {\em literal identity} is not a formal notion.
What we mean is as follows:
the diverse components of a computation device in $\+C^+$
are in one-one correspondance
with the diverse components of the associated EMA,
and this correspondance is an identity up to a change of perspective
(for instance, a ``physical" bi-infinite tape will be considered to be
identical to the mathematical set $\Z$ of integers).
\begin{remark}\label{rk:main}
1. This theorem is indeed a schema: one theorem per computation model.
We have proved it for a variety of usual sequential computation models
(cf. \cite{grigovalarcher_web}).
\\
2. As said above, the diverse instances of Theorem \ref{thm:main} 
are proved for slight extension $\+C^+$ of the usual computation models $\+C$.
In all cases, $\+C^+$ can be viewed as $\+C$ considered with different time units:
for any $k\geq 1$, a device $\+M$ in $\+C$
is seen as a device $\+M^{(k)}$ in $\+C^+$ in which 
one step of $\+M^{(k)}$ corresponds to $k$ successive steps of $\+M$
(or $< k$ successive steps in case the last of these steps has no
successor).
\\
3. Considering another presentation of $\+C^+$, one can also view it as $\+C$
in which some contingencies have been removed
(for instance, the read/write head will be able to scan a window of cells
instead of a single cell)
but the computational paradigm has been preserved:
local computation and a particular topology of data storage
for Turing machines,
indirect addressing of registers for random access machines, etc.
In our opinion, the classes $\+C^+$ are the {\em right ones} to carry
the diverse computation paradigms.
\\
4. In fact, contingencies can also be captured
by families of EMAs with more technical definitions
(cf. \cite{grigovalarcher_web}):
we loose the remarkable simplicity of the above families $\+E_{\+M,\+S}$.
\\
5. This theorem schema strengthens Gurevich's claim that ASMs constitute
the natural mathematical modelization of algorithms:
{\em EMAs (which are a variant of ASMs)
appear as the computation model which unifies
all usual sequential computation paradigms}.
\end{remark}
\noindent{\bf About the proof.}
No surprise, the proof of Theorem \ref{thm:main}
for a particular $\+C$ involves the particular
features of the class $\+C$.
Thus, the claim (point 5 in Remark \ref{rk:main})
that Theorem \ref{thm:main} is true
for extensions $\+C^+$ of every usual sequential computation model
$\+C$ cannot be proved but only be supported by proved instances
for a variety of classes $\+C$.

As for the common features to all such proofs, they come from an analysis
of what precludes positive solutions to questions (Q1) and (Q2). 
Let us list some of the difficulties which are met.
Some are easy to solve, other ones force to
adequately tailor the definition of ASMs (as that of EMAs)
and those of the usual computation models.
\medskip\\
(1)\
An ASM program mimicking the transition function $\delta$ of
a Turing machine is a description of $\delta$.
Since there are many distinct descriptions of the same $\delta$,
there are many ASMs which tightly simulate the same Turing machine.
Thus, surprisingly as it is, looking at this component
-- transition functions --,
ASMs are less abstract than Turing machines.
Somehow, there is an extra operational feature in ASMs:
the operational way to use $\delta$ is not part of the formalization
of Turing machines.
\\
This is why we modify ASMs to EMAs:
{\em Evolving Multialgebras}.
The notion of EMA is that of ASM in which the program
(a syntactic object) is replaced by a semantic object:
a (very simply definable) functional operating on the function sets
over the ASM domain.
It is then more natural to break the universe of
an ASM into its natural parts: this allows a very useful rudimentary
typing of elements and functions.
\medskip\\
(3)\
Again considering Turing machines,
an ASM simulates the tape by the set $\Z$ of all integers
and the moves of the head by the successor and predecessor
operations on $\Z$.
Terms in the ASM logical language allow to name the
$i$-th successor and the $i$-th predecessor.
Thus, we cannot avoid the ASM program to move the head
more than one cell left or right
unless we constrain terms in ASM programs to be of a simple form
(somewhat ``flat").
Which would put technicalities to any positive answer to question
(Q2).
This is why we consider slight extensions of the machine models
which allow the read/write head to scan a window of cells rather than
only one cell and to move in a window.
This is a kind of extra capability which is much like
allowing several tapes or several heads.
Though it does modify the model, it does preserves its core feature:
successive local actions.
\medskip\\
(4)\
For machine models having programs
like RAMs and SMM (Sch\"onhage  Storage ModiÞcation Machines),
there are two slight modifications.
First, allow bounded blocks of parallel and/or successive actions.
Second, remove the program and the program counter
in favor of a transition function
(much in the vein of Turing machines)
which, though operating on an infinite set
(the contents of the accumulator and of the addressed registers in the case of RAMs) is very simply definable
in terms of the original program.
Thus, we replace an operational item (the program) by a denotational
one (the transition function).
Again, though it does modify the model, it does preserves its core feature: indirect addressing (for RAMs),
dynamic storage topology (for Sch\" onhage pointer machines).
\medskip\\
{\bf EMAs versus ASMs.}
In our opinion, ASMs and EMAs are complementary models.
EMAs generalize any type of machine: it is the unification model.
As for ASMs, they are closer to programming.
Indeed, the functioning of a EMA
 goes through the iteration of a functional.
To program an EMA, we need to add some operational information
on how to use this functional and this leads back to a program,
hence to an ASM\ldots
Thus, ASMs are EMAs plus the instructions for using the functional:
ASMs refine EMAs (in the sense of software engineering)
and EMAs are a (more) abstract version of ASMs.

%
%
%
\section{From ASMs to EMA
s: the deterministic case}\label{s:from}
%
\subsection{How EMAs differ from ASMs}
\label{ss:differ}
%
We detail the diverse features which are peculiar to EMAs. 
\medskip\\
{\bf A functional in place of a program.}
As said in the introduction,
the main difference between evolving multialgebras and
Gurevich's ASMs is as follows:
{\em the program (i.e. a syntactic object) of an ASM
is replaced by a functional (i.e. a semantic object)}
which does exactly what the program tells to do.
Thus, an operational feature is removed.
\medskip\\
{\bf Multi-domains and multialgebras.}
The above modification leads to another very minor one,
really kind of ``semantic sugar":
{\em the universe of an ASM is broken into its natural constituents
and becomes a multi-domain}.
The reason for such multialgebras is that they make it possible to
type the symbols of the signature as functions (or elements)
between the diverse sets of the multi-domain.
\medskip\\
{\bf Multialgebra operations with values in products of domains.}
Set theoretically, a map $F:A\to B\times C$ is identified with the pair
of its component maps
$(F_B,F_C)$ where $F_B:A\to B$ and $F_C:A\to C$.
We do view such an $F$ as the pair $(F_B,F_C)$ plus a correlation
condition: {\em one cannot fire one of these two component maps without
firing the other one, and both have to be fired on the same argument}.
\\
We allow operations in the multialgebra to take values in products of domains.
The above condition leads to a notion of multiterms and a constraint in the definition of formulas associated to the signature of an EMA.
It is used in \S\ref{s:schonhage} to deal with Sch\"onhage machines.
\medskip\\
{\bf Halt/Fail and EMA
 status.}
In EMAs, the ASM program is replaced by the functional
which does exactly what the program tells to do.
There are still the questions:
\\- is the  functional to be applied or not on given arguments?
\\- if not, does it
    ``halts and accepts" or ``halts and rejects" or ``get stuck"?
\\
To deal with the three first alternatives, EMAs have a three valued
dynamic component: the status. Of course, there is no formal component carrying the information ``stuck".
\medskip\\
{\bf Inputs and ASMs.}
In most presentations, Gurevich does not give any formal status
to inputs (his paper \cite{dershowitzgurevich08} with Dershowitz
being an exception).
When dealing with question (Q2)
it turns out that it is important to give a formal status to inputs.
This is the case for EMA characterizations of machines having some
read-only tapes (e.g., finite automata).
We consider that {\em inputs appear in two ways:
\\- as values of some particular static symbols,
\\- as initial values of dynamic symbols.}
%
%
\subsection{Deterministic Evolving MultiAlgebras}
\label{ss:EMA}
%
\begin{definition}\label{def:types}
Let $n\geq1$ and $\+D=(D_i)_{i=1,\ldots,n}$ be a sequence
of $n$ non empty sets (which we call an $n$-multiset).
An $n$-sort type is a triple $(k,\alpha,\ell)$ where
$k\in\N$, $\ell\in\{1,\ldots,n\}$
and $\alpha$ is a map $\{1,\ldots,k\} \to \{1,\ldots,n\}$.
Its associated $\+D$-type $(k,\alpha,\ell)_{\+D}$
is the family of all partial functions
$D_{\alpha(1)}\times\ldots\times D_{\alpha(k)} \to D_\ell$.
A $\+D$-type is functional if $k\geq1$.
In case $k=0$, the $\+D$-type $(0,\emptyset,\ell)_{\+D}$
is the family of partial functions
$\{\emptyset\}\to D_\ell$, i.e. the set of ``partial elements" of $D_\ell$,
i.e. $D_\ell$ augmented with an ``undefined element".
\\
{\em Intuition: there are $k$ arguments,
$\alpha$ gives their types, and $\ell$ is the type of the range.}
\\
Typed ground terms and their types are defined in the obvious way.
\end{definition}
\noindent{\bf Multialgebras.}
The notion of multisort algebra is a direct extension to multiset domains
of the usual notion of algebra of partial functions on a unique domain.
\begin{definition}[Multialgebras]
Let $n\geq1$ and $\+S$ be an $n$-sort typed signature
containing function symbols $\varphi_1,\ldots, \varphi_p$.
An $\+S$-multialgebra $\+A$ is an $n$-multiset
 $\+D=(D_i)_{i=1,\ldots,n}$
endowed with partial functions $F_1,\ldots,F_p$ which interpret
the symbols $\varphi_i$'s (Care: arity $0$ symbols with type $D_i$
are interpreted by elements of $D_i$ but can also be undefined).
\\
If defined, the value, relative to $\+A$, of a ground $\+S$-term $t$
is denoted by  $\means{t}_{\+A}$ (it is an element of some $D_i$).
\end{definition}
\noindent{\bf Semialgebraic functionals.}
Semialgebraic functionals are those which can be described
by ASM programs.
They modify the interpretations in the multialgebra
of constant and functions symbols.
For function symbols, this modification affects the values
of only finitely many points in the domain.
These points and the associated new values of the argument
are given by ground $\+S$-terms.
As in ASMs programs, there is a disjunction of cases for the
choice of the affected points and their associated new values.

First, a convenient notion.
\begin{definition}[The $\oplus$ operation]
Let $F,G$ be partial functions $X_1\times\ldots\times X_k\to Y$ and
$Z\subseteq X_1\times\ldots\times X_k$.
We define the partial function $F\oplus_Z G$ as follows:
\begin{eqnarray*}
\dom(F\oplus_Z G) &=& (\dom(F)\setminus Z)\cup(\dom(G)\cap Z)
\\
(F\oplus_Z G)(\vec{x}) &=&
\left\{\begin{array}{ll}
F(\vec{x})&\textit{if\quad $\vec{x}\notin Z$}\\
G(\vec{x})&\textit{if\quad $\vec{x}\in Z$}
\end{array}\right.
\end{eqnarray*}
In case $p=0$, $F,G$ are ``partial elements" of $Y$
and $Z\subseteq\{\emptyset\}$ and
$F\oplus_Z G=F$ if $Z=\emptyset$ and
$F\oplus_Z G=G$ if $Z=\{\emptyset\}$.
\end{definition}
\begin{definition}[\bf Semialgebraic functionals]
\label{def:semialgebraic}
Let
\\\textbullet\ $\+D=(D_i)_{i=1,\ldots,n}$ be an $n$-multiset,
\\\textbullet\ $\+S$ be an $n$-sort typed signature
containing function symbols $\varphi_1,\ldots, \varphi_p$,
\\\textbullet\ $\+A$ be a multialgebra with signature
$\+S\setminus\{\varphi_1,\ldots, \varphi_p\}$ on $\+D$,
\\\textbullet\ $\+F_1,\ldots,\+F_p$ be the $\+D$-types associated to
$\varphi_1,\ldots, \varphi_p$,
\\\textbullet\ $m\in\{1,\ldots,p\}$ and $(k,\alpha,\ell)$ be the
$n$-sort type of $\varphi_m$. 
\\\textbullet\ $\+T_i$ be the family of ground $\+S$-terms of type
$D_i$,
\\
For any $p$-tuple of functions
$\vec{F}=(F_1,\ldots,F_p)\in\+F_1\times\ldots\times\+F_p$,
let us denote by $\+A(\vec{F})$
the multialgebra $\+A$ expanded to the signature $\+S$
in which the $\varphi_i$'s are interpreted by the $F_i$'s.
\\
A partial functional
$\Phi : \prod_{j=1,\ldots,p}\+F_j \longrightarrow \+F_m$
is $(\+S,\+A)$-semialgebraic if there exists a map
$\beta : \Bool^q \to \Pfin(\+T_{\alpha(1)}\times\ldots
                                        \times\+T_{\alpha(k)}
                                        \times\+T_{\ell})$
(where $\Pfin(X)$ is the family of finite subsets of $X$)
and ground $\+S$-terms $t_1,\ldots,t_q$, $t'_1\ldots,t'_q$
such that, for any $\vec{F}\in\+G_1\times\ldots\times\+G_p$,
$$\begin{array}{l}
\Phi(\vec{F})\mbox{ is defined if and only if}
\\
\quad\left\{\begin{array}{l}
(a)\ \ \mbox{all 
$\means{t_i}_{\+A(\vec{F})}$'s, $\means{t'_i}_{\+A(\vec{F})}$'s
are defined}
\\
(b)\ \ \forall (u_1,\ldots,u_k,v)\in \beta(\ldots,
	\means{t_i}_{\+A(\vec{F})}= \means{t'_i}_{\+A(\vec{F})},\ldots)
\mbox{ all $\means{u_j}_{\+A(\vec{F})}$'s are defined}
\\
(c)\ \ \forall (\vec{u},v), (\vec{w},z)\in \beta(\ldots,
	\means{t_i}_{\+A(\vec{F})}= \means{t'_i}_{\+A(\vec{F})},\ldots)
\quad\mbox{ $\means{u_j}_{\+A(\vec{F})} \neq
               \means{w_j}_{\+A(\vec{F})}$    for some $j$}
\end{array}\right.
\medskip\\
\Phi(\vec{F}) = F_m \oplus_Z G
\mbox{ where }\\
\quad Z=\{
(\means{u_1}_{\+A(\vec{F})},
\ldots, 
\means{u_k}_{\+A(\vec{F})}) \mid
\exists v\ (\vec{u},v) \in \beta(\ldots,
\means{t_i}_{\+A(\vec{F})}= \means{t'_i}_{\+A(\vec{F})},\ldots)\}
\\
\quad G=\{
(\means{u_1}_{\+A(\vec{F})},
\ldots, 
\means{u_k}_{\+A(\vec{F})},
\means{v}_{\+A(\vec{F})}) \mid
(\vec{u},v) \in \beta(\ldots,
\means{t_i}_{\+A(\vec{F})}= \means{t'_i}_{\+A(\vec{F})},\ldots)\}
\end{array}$$
The tuple $(\beta,t_1,\ldots,t_q,t'_1\ldots,t'_q)$
is called a presentation of $\Phi$.
\medskip\\
For $I\subseteq\{1,\dots,p\}$, a functional
$\Psi : \prod_{j=1,\ldots,p}\+F_j \longrightarrow \prod_{m\in I}\+F_m$
is $(\+S,\+A)$-semialgebraic if so are all its components.
\end{definition}
\begin{remark}
Condition (a) in Definition \ref{def:semialgebraic} insures that
all equality tests $t_i=t'_i$ can be achieved.
Conditions (b) and (c) insure that, in equality
$\Phi(\vec{F})=F_m \oplus_Z G$,
the finite set $Z$ can be computed and
$G$ is a functional graph.
\\
We do not require the $\means{v}_{\+A(\vec{F})}$'s to be defined
(i.e. $\dom(G)=Z$):
though this is incompatible with a call by value strategy, it makes sense
with a call by name strategy.
\end{remark}
\begin{definition}[\bf Deterministic EMAs]\label{def:detEMA}
A deterministic evolving multialgebra (EMA) is a tuple
$
\+A = (n;\
\+S_{\sta}, \+S^{\sta}_{\inp},
\+S^{\dyn}_{\inp}, \+S_{\dyn};\
               \+D;\    \+M_{\sta}, \+M_{\ini};\    \Phi)
$
consisting of the following items.
\begin{itemize}
\item
An $n$-multiset $\+D=(D_i)_{i=1,\ldots,n}$ such that
$D_n=\{\go, \acc, \rej\}$.
\\
{\em Intuition. Sets $D_1,\ldots,D_{n-1}$ are the $n-1$ different sorts of objects
and $D_n=\{\go, \acc, \rej\}$ is the set of possible statuses
of the (evolving) multialgebra during the run:
``go on", ``halt and accept", ``halt and reject".}
\item
Four disjoint $n$-sort typed finite signatures
$\+S_{\sta}, \+S^{\sta}_{\inp},
\+S^{\dyn}_{\inp}, \+S_{\dyn}$
and two structures $\+M_{\sta}, \+M_{\ini}$ with respective
signatures $\+S_{\sta}, \+S_{\dyn}$.
There is only one symbol $\fraks$ which involves the sort $n$ :
it is a constant of type $D_n$ in $\+S^{\dyn}_{\inp}$.
\\
{\em Intuition. $\+M_{\sta}$ is the static framework on $\+D$
which remains fixed during any run.
$\+S^{\sta}_{\inp}$ is the signature for the static part of the input:
its interpretation remains fixed (hence accessible) during a run.
$\+S^{\dyn}_{\inp}$ is the signature for the dynamic part of the input:
its interpretation can be modified (hence become inaccessible)
during a run.
$\+M_{\ini}$ initializes the part of the dynamic environment
which is not initialized by the input.
The interpretation of $\fraks$ represents the status of the multialgebra}.
\item
Let $\+S=\+S_{\sta}\cup\+S^{\sta}_{\inp}\cup\+S^{\dyn}_{\inp}\cup\+S_{\dyn}$.
$\Phi$ is a $(\+S,\+M_{\sta})$-semialgebraic partial functional
$$
\Phi  : \left( \prod_{\varphi\in \+S^{\sta}_{\inp}}  \+F_\varphi \right)
\times
\left(\{\go\} \times
 \prod_{\varphi\in (\+S_{\dyn}\cup\+S^{\dyn}_{\inp})
                              \setminus\{\fraks\}}  \+F_\varphi \right)
\quad\longrightarrow \quad
\prod_{\varphi\in \+S_{\dyn}\cup\+S^{\dyn}_{\inp}}  \+F_\varphi
$$
where $\+F_\varphi$ denotes the semantic type of the function symbol
$\varphi$.
In particular, $\Phi$ rules the evolution of the status.
The sole status which can be an argument of $\Phi$ is ``go":
a multialgebra with status ``acc" or ``rej" is halted
and does not evolve any more.
However, in the image of $\Phi$ the status can take any value.
\end{itemize}
A state of $\+A$ is any multialgebra on $\+D$ with signature
$\+S$ which expands $\+M_{\sta}$.
\end{definition}
\begin{definition}[\bf Runs of deterministic EMAs]
We keep the notations of Definition \ref{def:detEMA}.
A run of $\+A$ is a sequence $(\+M_t)_{t\in I}$ of states of $\+A$
such that
\\\textbullet\
$I$ is a finite or infinite non empty initial segment of $\N$,
\\\textbullet\
$\means{\theta}_{\+M_0} = \means{\theta}_{\+M_{\ini}}$
for all $\theta\in\+S_{\dyn}$,
\\\textbullet\
If $t\in I$ then
$\means{\theta}_{\+M_t} = \means{\theta}_{\+M_0}$
for all $\theta\in\+S^{\sta}_{\inp}$,
\\\textbullet\
If $t\in I$ then $t+1$ is in $I$ if and only if
$\means{\fraks}_{\+M_t}= \go$ and $\Phi$ is defined on
$(\means{\varphi}_{\+M}\})_{\varphi\in\+S \setminus \+S_{sta}}$,
\\\textbullet\
If $t+1\in I$ then 
$(\means{\theta}_{\+M_{t+1}})_{\theta\in\+S_{\dyn}\cup\+S^{\dyn}_{\inp}}=
\Phi((\means{\varphi}_{\+M_t}\})_{\varphi\in\+S \setminus \+S_{sta}})$.
\\
In particular, if  $\means{\fraks}_{\+M_0} \neq \go$ then $I=\{0\}$.
Also, if $t+1\in I$ then $\means{\fraks}_{\+M_t}=\go$.
\end{definition}

%
\section{Turing machines}
\label{s:Turing}
%
In order to {\em identify} Turing machines with a simple class of EMAs,
we introduce a slight variant of Turing machines,
which we call ``window Turing machines":
1)
the head is allowed to scan a small window instead of a single cell,
and to move inside a window in a single step,
2)
halting (be it accepting or rejecting) is not related to the current state
but to the current local configuration: the state plus the contents of the scanned window.
\begin{definition}\label{def:Turing}
A deterministic $k$-window $n$-tape (bi-infinite tapes)
Turing machine is a tuple
$
(n,k,\Sigma=\{\sigma_0,\ldots,\sigma_{s-1}\},
Q=\{q_0,\ldots,q_{r-1}\}, F^+,F^-, \delta, \omega_i,\mu_i)_{i=1,\ldots,n}
$
where, for $i=1,\ldots,n$,
\begin{itemize}
\item
$\Sigma$ and $Q$ are finite sets (the alphabet and the set of states),
\item
$F^+,F^-\subseteq Q\times\Sigma^{n(2k+1)}$
(accepting/rejecting final local configurations),
\item
$\delta : Q\times\Sigma^{n(2k+1)} \to Q$
(state transition),
\item
$\tau_i : Q\times\Sigma^{n(2k+1)} \to \Sigma^{n(2k+1)}$
(read/write on tape $i$),
\item
$\mu_i : Q \times\Sigma^{n(2k+1)} \to\{-k,\ldots,-1,0,1,\ldots,k\}$
(move on tape $i$).
\end{itemize}
On each tape, the head scans the cell on which it is positioned
and the $k$ cells to the left and the $k$ cells to the right,
a total of $2k+1$ cells.
The argument of type $\Sigma^{n(2k+1)}$ in $\delta, \omega_i, \mu_i$
is the contents of the $n(2k+1)$ cells scanned on the $n$ tapes.
The effect of a transition is
to change the state according to $\delta$,
to modify the contents of the scanned cells of tape $i$
according to $\omega_i$ and to move its head according to $\mu_i $.
\\
The notions of run, halt, acceptance and rejection are defined as usual. 
\end{definition}
\begin{remark}
Usual deterministic $n$-tape Turing machines are the
$1$-window ones.
\end{remark}
\begin{definition}[\bf The class of EMAs for Turing machines]
\label{def:C_T}
We denote by $\+C^{(n)}_{\textit{wT}}$ the class of EMAs
$
\+A = (n+3;\
\+S_{\sta}, \+S^{\sta}_{\inp},
\+S^{\dyn}_{\inp}, \+S_{\dyn};\
               \+D;\    \+M_{\sta}, \+M_{\ini};\    \Phi)
$
which
satisfy the following conditions for some $r,s\in\N$
(for clarity, we abusively denote by the same letter static constant symbols
and the elements which interpret them in the structure $\+D$).
\medskip\\
(1) 
The multidomain of $\+A$ is
$
\+D=(\Z^{(1)},\ldots,\Z^{(n)}, Q, \Sigma, \frakS)
$
where the $\Z^{(i)}$'s are fixed pairwise disjoint copies of $\Z$
(for instance, $\Z^{(i)}=\Z\times\{i\}$),
$Q,\Sigma$ are finite sets with $r, s$ elements respectively,
and $\frakS = \{\go,\acc,\rej\}$.
\medskip\\
(2) 
The static framework signature $\+S_{\sta}$ contains
$r$ constants $q_0,\ldots,q_{r-1}$ of type $Q$,
$s$ constants $\sigma_0,\ldots,\sigma_{s-1}$ of type $\Sigma$
and three constants $\go,\acc,\rej$ of type $\frakS $
which are interpreted in the obvious way in $\+M_{\sta}$.
It also contains, for each $i=1,\ldots,n$,
two unary functions symbols $\suc^{(i)},\pred^{(i)}$
of type $\Z^{(i)} \to \Z^{(i)}$ which are interpreted in $\+M_{\sta}$
as the successor and predecessor functions in $\Z^{(i)}$.
\medskip\\
(3) 
The signature $\+S^{\sta}_{\inp}$ is empty.
\medskip\\
(4) 
The signature $\+S_{\dyn}$
(for the dynamic environment non initialized by the input) contains,
for each $i=1,\ldots,n$, one constant $\pos^{(i)}$ of type $\Z^{(i)}$
one constant $q$ of type $Q$,
and one constant $\fraks$ of type $\frakS$,
which are respectively interpreted in $\+M_{\ini}$ as $0$,
$q_0$ and $\go$.
\medskip\\
(5) 
The signature $\+S^{\dyn}_{\inp}$
(for the dynamic environment initialized by the input) contains,
for each $i=1,\ldots,n$, one unary function $c^{(i)}$ of type
$\Z^{(i)}\to\Sigma$.
\medskip\\
Thus, the EMAs in $\+C^{(n)}_{\textit{wT}}$ are defined as those
having particular signature, multidomain, static framework
and initialization of some dynamic symbols
with no condition on the functional $\Phi$
(other than its semialgebraicity).
\end{definition}
\begin{theorem}[\bf EMA representation theorem for Turing machines]
\label{thm:TuringABRIDGED}$ \\$
Any deterministic $n$-tape window Turing machine is
{\em literally identical} to some EMA in the
class $\+C^{(n)}_{\textit{wT}}$.
Conversely, any EMA in $\+C^{(n)}_{\textit{wT}}$
is {\em literally identical}
to some deterministic $n$-tape window Turing machine.
\end{theorem}
\begin{proof}
The argument is based on the following literal identifications
between the components of a Turing machine (TM) and the
interpretations of symbols of the EMA signature:
\begin{enumerate}
\item
(TM) $i$-th tape and the way the read/write head moves on it.
\\
(EMA) the copy $\Z^{(i)}$ of $\Z$ structured as
$\langle \Z^{(i)},\suc^{(i)}, \pred^{(i)} \rangle$.
\item
(TM) diverse states and letters.
\\
(EMA) interpretations of the static symbols $q_0,\ldots,q_{r-1}$
and $\sigma_0,\ldots,\sigma_{s-1}$.
\item
(TM) current state, positions of the $n$ heads
and contents of the $n$ tapes.
\\
(EMA) current interpretations of the dynamic symbols
$q$, $pos^{(i)}$, $c^{(i)}$.
\item
(TM) non final or final accepting/rejecting character of the current state.
\\
(EMA) current interpretation of the dynamic symbol $\fraks$.
\item
(TM) transition function.
\\
(EMA) semialgebraic functional.
\item
(TM) initial configuration.
\\
(EMA) interpretations of the $c_i$'s in the initial multialgebra
and of $\+S^{\dyn}_{\dyn}$ in $\+M_{\ini}$.
\end{enumerate}
The non trivial identifications are those of points 4 and 5.
\\
Keeping the notations of Definition \ref{def:semialgebraic}, let
$(\beta_\varphi,t_{1,\varphi},\ldots,t_{q_\varphi,\varphi},
 t'_{1,\varphi}\ldots,
 t'_{q_\varphi,\varphi})_{\varphi\in\+S^{\inte}_{\dyn}}$
be a presentation of the semialgebraic functional $\Phi$
of an EMA:
$$\beta_\varphi : \Bool^{q_\varphi}
                                    \to \Pfin(\+T_{\alpha_\varphi(1)}
                                        \times\ldots
                                        \times\+T_{\alpha_\varphi(k_\varphi)}
                                        \times\+T_{\ell_\varphi})
$$
Observe that terms of type $\Z^{(j)}$ are of the form
$\xi_1(\xi_2(\ldots))(\pos^{(j)})$
where the $\xi_k$'s are $\suc^{(j)}$ or $\pred^{(j)}$.
Let $k$ be the maximum value of the $|\xi_1(\xi_2(\ldots))(0)|$
for all terms of type some $\Z^{(j)}$ which is among the
$t_{i,\varphi},t'_{i,\varphi}$ or among the finite sets given by
the $\beta_\varphi$'s.

First, let us look at the equalities $t_{i,\varphi}=t'_{i,\varphi}$
which govern the domain of $\Phi$.
\\
\textbullet\
If $t_{i,\varphi},t'_{i,\varphi}$ have type $\Z^{(j)}$
then, as said above,
they are of the form $\xi_1(\xi_2(\ldots))(\pos^{(j)})$.
Hence any equality $t_{i,\varphi}=t'_{i,\varphi}$ is trivially
true or false independently of the current value of $\pos^{(j)}$.
\\
If $t_{i,\varphi},t'_{i,\varphi}$ have type $\frakS$ 
then they are of the form $\fraks$ or $\go, \acc, \rej$.
Since $\Phi$ and $\beta$ are restricted to values where
$\fraks=\go$, all possible equalities are trivial.
\\
Thus, we can suppose that there is no term with type
$\Z^{(j)}$ or $\frakS$ among the $t_{i,\varphi},t'_{i,\varphi}$'s.
\\
\textbullet\
If $t_{i,\varphi},t'_{i,\varphi}$ have type $Q$
then they are of the form $q$ or $q_j$ ($j=0,\ldots,r-1$).
Since any equality $q_j=q_k$ is trivially true or false,
we can suppose that there is at most one equality between terms
of type $Q$ and that it is of the form $q=q_j$.
\\
\textbullet\
If $t_{i,\varphi},t'_{i,\varphi}$ have type $\Sigma$
then they are of the form
$c^{(j)}(\xi_1(\xi_2(\ldots))(\pos^{(j)}))$
where the $\xi_k$'s are $\suc^{(j)}$ or $\pred^{(j)}$.
The equalities between terms of type $\Sigma$ are all comparisons of letters among the values of 
$c^{(1)}(-k),\ldots,c^{(1)}(k)$,\ldots, $c^{(n)}(-k),\ldots,c^{(n)}(k)$
where $k$ is defined above.

This shows that the values of $\Phi$ depend solely on the value of
$q$ and those of the $c^{(j)}(\pos^{(j)}+i)$'s
for $j=1,\ldots,n$ and $i=-k,\ldots,k$).
This is exactly to say that what matters is the current state
and the current letters in the $n$ windows
of diameter $2k+1$ centered at the positions of the $n$ heads.
Otherwise said, the tuple of arguments of the functional $\Phi$
is literally identical to the current values of the state plus the contents
of the windows, that is a tuple in $Q \times \Sigma^{n(2k+1)}$.
\medskip\\
Let us look at the image of $\Phi$ which is given through finite
families of tuples of terms given by the $\beta_\varphi$'s.
Since the only terms of type $Q$ are $q$ and the $q_i$'s.
Thus, $\Phi$ can leave the dynamic symbol $q$ unchanged
or modify it to any value.
The same is valid for the dynamic symbol $\fraks$
(using what is said above about the domain of $\Phi$, this
proves the non easy direction of point 4). 
\\
Terms of type $\Sigma$ name the contents of some $c^{(j)}$
at positions which are at distance $\leq k$ of the position of the $j$-th head.
Thus $\Phi$ can modify the values of the $c^{(j)}$ in the windows
around the positions of the heads.
\\
Terms of type $\Z^{(j)}$ name an integer at distance $k$ of the position of the $j$-th head.
Thus $\Phi$ can move any head left or right of at most $k$ cells.
This proves the non easy direction of point 5.
Thus, an EMA in $\+C^{(n)}_T$ is literally identical
to some window Turing machine.
The converse is proved in a similar (much easier) way.
\end{proof}
\begin{remark}
A slight variation in the EMA model can have strong effect.
For instance, suppose we add a constant $0$ to the static signature
and interpret it as $0$ in the structure $\+M_{\sta}$. 
Then we get window Turing machines in which the head can jump
to cell $0$.
\end{remark}
%
%
%
%
%
\section{Random access machines}
\label{s:RAM}
In order to {\em identify} RAMs with a simple class of EMAs,
we introduce a slight variant of RAMs,
which we call ``transition RAM" (TRAM):
1)
a bounded number of registers can be modified in one step,
2)
it can test for equality to $0$ and equality between combinations
(via the fixed set of operations on $\N$) of the contents of the
addressed registers,
3)
the program is replaced by a transition function. Though this function
operates on an infinite domain, it is finitarily defined via ground terms.
\begin{definition}[$n$-transition RAMs]
Let $f_1,\ldots,f_p$ operations on non negative integers,
A $n$-transition RAM ($n$-TRAM)
with operations $f_1,\ldots,f_p$ is a tuple
$$
(n,k, Q=\{q_0,\ldots,q_{r-1}\},
F^+,F^-, \delta, \rho_i, \tau_{i,j})_{i=1,\ldots,n,\
j=1,\ldots,k}
$$
where
\begin{itemize}
\item
$n$ is the number of distinguished registers,
\item
$\Sigma$ and $Q$ are finite sets (the alphabet and the set of states),
\item
$F^+,F^-\subseteq Q\times\Bool^p$
(accepting/rejecting final local configurations),
\item
$\delta : Q\times \Bool^p \to Q$
(state transition),
\item
$\rho_i : Q\times \Bool^p \to T$
(modification of register $i$)
for $i=1,\ldots, n$,
where $T$ is a finite family of terms built with the operations $f_1,\ldots,f_p$ and $n(1+k)$ constants
(representing the contents of the addressed registers),
\item
$\tau_{i,j} : Q\times \Bool^p \to T$
(modification of the register addressed through an iteration of $j$ successive addressing, starting with register $i$),
for $i=1,\ldots, n$, $j=1,\ldots,k$.
\end{itemize}
At any time the $n$-TRAM accesses registers $1,\ldots, n$
and the registers addressed addressed through at most $k$ iterated
addressing by these registers.
The $p=n(1+k)(1+\frac{n(1+k)-1}{2})$ Boolean arguments
in the $\delta,\rho_i,\tau_i$'s
test equalities or equalities to $0$ of the contents of the $n(1+k)$
adressed registers. 
Map $\delta$ tells how the state is modified.
Maps $\rho_i,\tau_{i,j}$'s tell how the contents of the accessed registers
are modified.
\\
The notions of run, halt, acceptance and rejection are defined
in the usual way. 
\end{definition}
\begin{definition}[\bf The class of EMAs for TRAMS]\label{def:C_RAM}
Let $f_1,\ldots,f_p$ operations on non negative integers.
We denote by $\+C^{(n)}_{\text{TRAM}}$ the class of EMAs
$\+A$
which satisfy the following conditions.
\medskip\\
(1) 
$\+A$ has $4$ sorts and its multidomain is
$
\+D=(\N, \N^{\adr}, Q, \frakS)
$
where $\N^{\adr} $ is a copy of $\N$,
$Q$ is a finite set with $r$ elements,
and $\frakS = \{\go,\acc,\rej\}$.
\medskip\\
(2) 
The signature $\+S_{\sta}$ (for the static framework) contains
$n+r+3$ constants: $1,\dots,n$ of type $\N$,
$q_0,\ldots,q_{r-1}$ of type $Q$,
``go", ``acc", ``rej" of type $\frakS $,
and $n+1$ unary function symbols {\sl cast}
of type $\N\to N^{\adr}$,
and, for each $i=1,\ldots,n$, $f_i$ of type $\N^{k_i}\to\N$.
Their interpretations in $\+M_{\sta}$  are as follows:
i)~$f_i$ is interpreted as the given operation on $\N$,
ii)~the $\cast$ function is interpreted as the identity from $\N$
to its copy $N^{\adr}$,
iii)~$1,\ldots,n$, the $q_i$'s and ``go", ``acc", ``rej"
are interpreted in the obvious way.
\medskip\\
(3) 
The signature $\+S^{\sta}_{\inp}$ is empty.
\medskip\\
(4) 
The signature $\+S_{\dyn}$ contains
two constants $q, \fraks$ of types $Q$ and $\frakS$.
Their interpretations in $\+M_{\ini}$ are $q_0$ and ``go".
\medskip\\
(5) 
The signature $\+S^{\dyn}_{\inp}$ contains
one unary function $c$ of type $\N^{\adr}\to\N$.
%
\medskip \\
Thus, the EMAs in $\+C^{(n)}_{\text{TRAM}}$ are defined as those
having particular signature, multidomain, static framework
and initialization of some dynamic symbols
with no condition on the functional $\Phi$
(other than its semialgebraicity).
\end{definition}
\begin{theorem}[\bf EMA representation theorem for TRAMs]
\label{thm:RAM}$ \\$
Any $n$-TRAM is {\em literally identical} to some EMA
in the class $\+C^{(n)}_{\text{TRAM}}$.
Conversely, any EMA
 in $\+C^{(n)}_{\text{TRAM}}$
is {\em literally identical} to some $n$-TRAM.
\end{theorem}
\begin{proof}
Analogous to the proof of Theorem \ref{thm:TuringABRIDGED}.
\end{proof}
%
%
%
%
%
%
%
\section{Other models}
\label{s:automata}
%
Similar results can be proved with finite atomata, stack automata
Sch\"onhage machines.

\medskip

Let us mention an interesting feature occurring in the EMA modelization
of Sch\"onhage Storage Modification Machines (SMM) which illustrates what has been said in \S\ref{ss:differ} about operations with values in products of domains.
The tape of an SMM is a dynamic graph which may grow or loose nodes.
To manage the current set of nodes of this graph-tape, it is convenient
to introduce the following items:
\\\textbullet\
Among the sets of the multi-domain $\+D$, there is an infinite set $X$ 
(where all nodes are taken) and the set $\Pfin(X)$ of finite subsets of $X$. There is no structure on $X$ nor on $\Pfin(X)$.
\\\textbullet\
In the signature $\+S_{\dyn}$, there is a constant symbol $U$ of type
$\Pfin(X)$ (it tells which nodes are in the current graph-tape).
\\\textbullet\
In the signature $\+S_{\sta}$, there is a function symbol $\textit{new}$
with type $\Pfin(X) \to X\times \Pfin(X)$. It is interpreted as a choice function $A \mapsto (a,A\cup\{a\})$ which picks in $X$ a point
outside $A$, i.e. such that $a\notin A$.

To add a new node to the graph tape, we apply $\textit{new}$ to $U$.
The constraint that both components of $\textit{new}$ have to be fired
simultaneously and on the same argument insures that when a new node
is picked, it is automatically added to (the interpretation) of $U$ with no
condition on the functional $\Phi$. 

%
\section{Uniformly bounded non determinism}
\label{s:boundednondet}
%
Uniformly bounded non determinism allows at each step at most $k$ choices where $k$ is some fixed constant independent of the step.
EMAs with `such non determinism are defined as are
deterministic EMAs with the following modification:
{\em replace the semialgebraic functional $\Phi$ by finitely many
such functionals.}
%
%
All litteral identity results mentioned in the previous sections
extend easily to the non deterministic cases.
%
\section{External non determinism}
%
We now deal with a more powerful kind of non determinism:
that given by external choices which may be done during the run.
This is the action of Gurevich's ``Choose" instruction.
To deal with such an ``external non determinism", we enrich EMAs
with a fifth signature: the ``external dynamic" signature
$\+S_{\ext}$.
We illustrate this notion with the example of Chomsky type $0$ grammars.
\begin{definition}
A grammar is a finite set of rules $(u_i,v_i)_{i=1,\ldots,n}$
where the $u_i,v_i$'s are words in an alphabet $\Sigma$.
The associated relation
$R\subseteq \Sigma^{\star} \times \Sigma^{\star} $
is defined as follows:
a pair $(U,V)$ is in $R$ if and only if there exists a finite sequence
$U=U_0,\ldots,U_k=V$ such that, for all $j<k$ there exists words
$P,S$ and some $i=1,\ldots,n$ such that
$U_j=Pu_iS$ and $U_{j+1}=Pv_iS$.
\end{definition}
\begin{definition}
We denote by $\+C_{\gra}$ the class of non deterministic EMAs
$$
\+A = (3;\
\+S_{\sta}, \+S^{\sta}_{\inp},
\+S^{\dyn}_{\inp}, \+S_{\dyn}, \+S_{\ext};\
               \+D;\    \+M_{\sta}, \+M_{\ini};\    \Phi)
$$
which satisfy the following conditions.
\medskip\\
(1) 
$\+A$ has $3$ sorts and its multidomain is
$
\+D=(\N, \Sigma^*,\frakS)
$
where $\Sigma$ is a finite set.
\medskip\\
(2) 
The signature $\+S_{\sta}$ (for the static framework) contains
finitely many binary function symbols $\textit{subst}_i$,
$i=1,\ldots,n$ of type $\N\times\Sigma^*\to \Sigma^*$.
There is some family $(u_i,v_i)_{i=1,\ldots,n}$ of pairs of words such
that the interpretation in $\+M_{\sta}$ (the static framework)
of $\textit{subst}_i$ is the function which acts on a pair $(p,U)$
as follows:
if $U$ contains the factor $u_i$ in position $p$
then it is replaced by $v_i$,
else $U$ is not modified.
\medskip\\
(3) 
The signatures $\+S^{\sta}_{\inp}$ and $\+S_{\dyn}$ are empty.
\medskip\\
(4) 
The signature $\+S^{\dyn}_{\inp}$ contains one constant $w$
of type $\Sigma^*$.
\medskip\\
(5) 
The signature $\+S_{\ext}$ (the external dynamic environment)
contains one constant $Choose$ of type $\N$.
Its interpretation during the run is given as an external action:
its value changes at each step.
%
\medskip \\
Thus, the EMAs in $\+C^{(n)}_{\gra}$ are defined as those
having particular signature, multidomain, static framework
and initialization of some dynamic symbols
with no condition on the functional $\Phi$
(other than its semialgebraicity).
\end{definition}
Using the fact that iteration of substitutions is also a substitution,
one can prove :

\begin{theorem}
Any grammar is {\em literally identical}
to some EMA in the class $\+C_{\gra}$.
Conversely, any EMA in $\+C_{\gra}$
is {\em literally identical} to some grammar.
\end{theorem}


\end{document}